\documentclass[sigconf]{acmart}

\copyrightyear{2026}
\acmYear{2026}
\setcopyright{cc}
\setcctype{by}
\acmConference[ICSE-SEIP '26]{2026 IEEE/ACM 48th International Conference on Software Engineering}{April 12--18, 2026}{Rio de Janeiro, Brazil}
\acmBooktitle{2026 IEEE/ACM 48th International Conference on Software Engineering (ICSE-SEIP '26), April 12--18, 2026, Rio de Janeiro, Brazil}
\acmPrice{}
\acmDOI{10.1145/3786583.3786900}
\acmISBN{979-8-4007-2426-8/2026/04}

\begin{document}

\title{Scaling Mobile Chaos Testing with AI-Driven Test Execution}

\author{Juan Marcano}
\authornote{These authors contributed equally to this work.}
\email{marcano@uber.com}
\affiliation{%
  \institution{Uber Technologies, Inc.}
  \city{Sunnyvale}
  \state{CA}
  \country{USA}
}

\author{Ashish Samant}
\authornotemark[1]
\email{asamant@uber.com}
\affiliation{%
  \institution{Uber Technologies, Inc.}
  \city{Sunnyvale}
  \state{CA}
  \country{USA}
}

\author{Kai Song}
\authornotemark[1]
\email{kaisong@uber.com}
\affiliation{%
  \institution{Uber Technologies, Inc.}
  \city{Sunnyvale}
  \state{CA}
  \country{USA}
}

\author{Lingchao Chen}
\authornotemark[1]
\email{lingchao@uber.com}
\affiliation{%
  \institution{Uber Technologies, Inc.}
  \city{Sunnyvale}
  \state{CA}
  \country{USA}
}

\author{Kaelan Mikowicz}
\authornotemark[1]
\email{kaelan@uber.com}
\affiliation{%
  \institution{Uber Technologies, Inc.}
  \city{Sunnyvale}
  \state{CA}
  \country{USA}
}

\author{Tim Smyth}
\authornotemark[1]
\email{smyth@uber.com}
\affiliation{%
  \institution{Uber Technologies, Inc.}
  \city{Sunnyvale}
  \state{CA}
  \country{USA}
}

\author{Mengdie Zhang}
\email{mengdiez@uber.com}
\affiliation{%
  \institution{Uber Technologies, Inc.}
  \city{Sunnyvale}
  \state{CA}
  \country{USA}
}

\author{Ali Zamani}
\email{azamani@uber.com}
\affiliation{%
  \institution{Uber Technologies, Inc.}
  \city{Sunnyvale}
  \state{CA}
  \country{USA}
}

\author{Arturo Bravo Rovirosa}
\email{arturo.bravo@uber.com}
\affiliation{%
  \institution{Uber Technologies, Inc.}
  \city{Sunnyvale}
  \state{CA}
  \country{USA}
}

\author{Sowjanya Puligadda}
\email{sowjanyap@uber.com}
\affiliation{%
  \institution{Uber Technologies, Inc.}
  \city{Sunnyvale}
  \state{CA}
  \country{USA}
}

\author{Srikanth Prodduturi}
\authornote{Senior advisors leading Uber's Failover Architecture.}
\email{sproddut@uber.com}
\affiliation{%
  \institution{Uber Technologies, Inc.}
  \city{Sunnyvale}
  \state{CA}
  \country{USA}
}

\author{Mayank Bansal}
\authornotemark[2]
\email{mbansal@uber.com}
\affiliation{%
  \institution{Uber Technologies, Inc.}
  \city{Sunnyvale}
  \state{CA}
  \country{USA}
}

\renewcommand{\shortauthors}{J. Marcano et al.}

\begin{abstract}
Mobile applications in large-scale distributed systems are susceptible to backend service failures, yet traditional chaos engineering approaches cannot scale mobile testing due to the combinatorial explosion of flows, locations, and failure scenarios that need validation. We present an automated mobile chaos testing system that integrates DragonCrawl, an LLM-based mobile testing platform, with uHavoc, a service-level fault injection system. The key insight is that adaptive AI-driven test execution can navigate mobile applications under degraded backend conditions, eliminating the need to manually write test cases for each combination of user flow, city, and failure type. 

Since Q1 2024, our system has executed over 180,000 automated chaos tests across 47 critical flows in Uber's Rider, Driver, and Eats applications, representing approximately 39,000 hours of manual testing effort that would be impractical at this scale. We identified 23 resilience risks, with 70\% being architectural dependency violations where non-critical service failures degraded core user flows. Twelve issues were severe enough to prevent trip requests or food orders. Two caused application crashes detectable only through mobile chaos testing, not backend testing alone. Automated root cause analysis reduced debugging time from hours to minutes, achieving 88\% precision@5 in attributing mobile failures to specific backend services. 

This paper presents the system design, evaluates its performance under fault injection (maintaining 99\% test reliability), and reports operational experience demonstrating that continuous mobile resilience validation is achievable at production scale.

\end{abstract}



\maketitle


\section{Introduction}

Mobile applications in large-scale distributed systems face a testing paradox: the more complex the backend becomes, the harder it is to validate how failures affect end users. At Uber, a single trip request flows through approximately 500 services with over 30,000 RPC calls. When non-critical services fail, which is inevitable, we need to verify that critical user flows remain unaffected. Traditional mobile testing cannot solve this problem at our scale.

The challenge is fundamentally one of combinatorial explosion. We have 61 critical user flows that behave differently across cities and respond differently to various backend failures. Comprehensive testing would require tens of thousands of individual test cases, each maintained separately as the application evolves. Manual testing is impractical at this scale. Automated UI tests break with every interface change. Chaos engineering works well for backend services but cannot capture how failures manifest in mobile user experiences. No existing approach bridges this gap.

We solved this by recognizing that two seemingly unrelated capabilities, AI-driven mobile test execution and systematic fault injection, could address the problem together if properly integrated. DragonCrawl, our LLM-based mobile testing system, maintains 99\% reliability despite UI changes by understanding screens through natural language rather than brittle selectors. uHavoc, our service-level fault injection platform, systematically injects backend failures with production-safe controls. Individually, neither system could validate mobile resilience at scale. Combined, they eliminate the need for manually writing test cases for each combination of flow, location, and failure scenario.

The integration required solving several non-trivial problems: coordinating fault injection timing with adaptive mobile test execution, attributing mobile failures to specific backend services across complex distributed traces, and operating safely at production scale. The resulting system has validated a core architectural principle: non-critical service failures should not cascade to impact core business functionality. Since Q1 2024, we have identified 23 violations of this principle—issues that would have become production incidents without proactive detection.

This paper makes the following contributions:
\begin{itemize}
\item An approach for integrating adaptive mobile testing with systematic fault injection that addresses the combinatorial explosion problem in mobile chaos testing
\item Implementation details of deploying this system at production scale, including automated root cause analysis and safety mechanisms
\item Quantitative evaluation demonstrating 99\% test reliability under fault injection and 88\% precision@5 for automated root cause analysis
\item Operational experience from 180,000+ test executions, including lessons learned and replication guidelines
\end{itemize}

Section 2 provides background on chaos engineering and describes Uber's operational complexity. Section 3 presents our system design. Section 4 discusses implementation challenges. Section 5 evaluates performance. Section 6 reports operational results. Section 7 surveys related work. Section 8 acknowledges limitations. Section 9 concludes with our takeaways.

\section{Background and Motivation}

\subsection{Chaos Engineering}

Chaos engineering emerged from a practical problem: distributed systems fail in unpredictable ways, and the only reliable method to validate resilience is to break things deliberately. Netflix pioneered this approach with Chaos Monkey, which randomly terminates production instances to verify that services can withstand instance failures~\cite{basiri2019automating}. The field has since evolved into sophisticated platforms that orchestrate complex failure scenarios while maintaining strict safety controls~\cite{rosenthal_book}.

Modern chaos engineering rests on several operational principles. Teams formulate specific hypotheses about system behavior under failure conditions rather than injecting random faults. Testing occurs in production or production-like environments because staging cannot replicate the complexity and interaction patterns of real systems. Automated safety mechanisms prevent experiments from exceeding acceptable blast radius. Validation happens continuously rather than only during major releases~\cite{basiri2016chaos}.

However, existing chaos engineering approaches focus almost exclusively on backend service resilience. Platforms like FIT~\cite{netflix_fit} and Gremlin excel at injecting failures into distributed systems, but they measure success through backend metrics: response times, error rates, circuit breaker activations. These systems cannot capture how backend failures manifest in mobile user experiences, where degradation may be subtle: a missing discount offer, a delayed screen transition, or a fallback UI state that confuses users.

\subsection{Mobile Testing Challenges}

Mobile testing has progressed from purely manual approaches to sophisticated automation frameworks, but fundamental scalability problems remain. Traditional automation relies on UI testing tools like Appium, Espresso, and XCUITest, which locate screen elements and execute predetermined action sequences~\cite{model_based_mobile}. This approach works for stable applications with predictable interfaces, but breaks down in dynamic production environments.

The core problem is brittleness. Mobile applications change constantly: feature flags alter behavior, A/B tests modify layouts, product updates redesign screens. Traditional automated tests depend on stable element identifiers and fixed UI hierarchies. When interfaces change, tests break, requiring manual updates that can exceed the maintenance value of automation itself~\cite{neto2021survey}. At Uber's scale, this maintenance burden becomes prohibitive.

Flow complexity compounds the problem. Modern mobile applications support numerous critical user journeys with city-specific variations and regulatory differences. Testing all combinations under various conditions quickly becomes unmanageable. A trip booking flow in San Francisco differs from Tokyo, which differs from São Paulo. Each variation requires separate test maintenance.

Testing environments introduce another gap. Mobile applications interact with complex backend systems that cannot be fully replicated outside production. Real user behavior patterns, data distributions, and failure modes differ significantly from test scenarios~\cite{automated_mobile_generation}. This means resilience issues often remain hidden until they impact actual users.

Recent AI-based testing approaches show promise. Large language models can interpret mobile screens through natural language understanding rather than brittle element selectors, enabling test automation that adapts to UI changes~\cite{wang2021vet}. However, these techniques have focused on functional testing under normal conditions. Applying them to resilience scenarios, where applications must continue functioning despite backend failures, remains unexplored.

\subsection{The Challenge of Mobile Resilience Testing}

Mobile applications at Uber serve as the critical interface for our core business flows. A single user request, such as booking a ride, involves coordination between hundreds of backend microservices including user authentication, location services, pricing, driver matching, payment processing, and real-time communications. In a typical trip flow test, the request flows through around 500 services, 100s in depth, and over 30,000 RPC calls. Out of many user experiences, there are 47 considered core business flows. Among these services, when any non-critical services fail or degrade, the mobile application must handle the failure gracefully to maintain user experience.

Traditional mobile testing approaches face several limitations when executing tests with fault injection:

\textbf{Manual Testing Limitations:} Most mobile testing is conducted manually, focusing on "happy path" scenarios under ideal conditions~\cite{model_based_mobile,neto2021survey,automated_mobile_generation}. Testing failure scenarios manually is time-consuming and difficult to scale across the breadth of possible service failures and user flows~\cite{model_based_mobile,neto2021survey,wang2021vet}.

\textbf{Limited Test Coverage:} Even with automated UI testing frameworks, achieving comprehensive coverage of critical user flows under failure conditions requires significant engineering effort to write and maintain tests for each failure scenario, and at Uber, even for every city. At Uber, we have 61 critical user flows which may behave differently in every city and certainly behave differently for every fault injection scenario. To illustrate the scale of this challenge, consider a simplified scenario: testing 61 flows across 50 cities with 7 fault injection scenarios would require over 20,000 individual test cases (61 x 50 cities x 7). This represents a classic combinatorial explosion problem: As we add more flows, cities, and failure scenarios, the number of required tests grows exponentially. In this scenario, adding just one new city would require 427 additional tests (61 flows × 7 scenarios), while adding a single new fault injection scenario would require 3,050 more tests (61 flows × 50 cities). This exponential growth makes traditional testing approaches fundamentally unscalable.

\textbf{Production Validation Gap:} Traditional chaos testing involves randomly injecting failures to verify overall system resilience. Although effective, this may cause serious Uber brand impacts, considering an Uber rider could be stranded on the street due to an ongoing chaos test. Uber's business cannot afford to tolerate risks like this. As a result, testing in staging environments becomes our only option, but it cannot fully replicate the complexity and unpredictability of production systems, creating a gap between test coverage and real-world resilience.

The core issue underlying these limitations is combinatorial explosion: the exponential growth in required test cases as we increase the dimensions of testing (flows × cities × failure scenarios). Traditional approaches that rely on manually writing individual test cases for each combination become mathematically intractable at Uber's scale. DragonCrawl addresses this fundamental scalability challenge by dynamically analyzing mobile application screens and intelligently deciding what actions to take in real-time. Rather than requiring engineers to hardcode tests for every combination of language, location, and fault-injection scenario, DragonCrawl adapts automatically to whatever screens and UI states it encounters, eliminating the need for pre-written test scripts across these multiple dimensions.

To address these challenges, our system is designed with the following requirements:
\vspace{-0.5\baselineskip}
\begin{enumerate}
\item Automatically execute mobile tests covering critical user flows, even in the presence of unexpected delays, screens, or UI variations introduced by fault injection.
\item Systematically inject backend service failures during test execution to expose resilience issues in a controlled and repeatable manner.
\item Scale testing across the breadth of Uber's mobile applications by dynamically analyzing screens and intelligently deciding actions in real-time, rather than requiring engineers to hardcode test combinations for different languages, locations, and fault-injection scenarios.
\item Provide actionable insights that help engineers prioritize and fix weaknesses in mobile application resilience.
\end{enumerate}

\subsection{Existing Solutions at Uber}

\textbf{uHavoc Platform:} The uHavoc Platform is Uber's service-level fault injection system~\cite{filibuster_paper}. It enables fault injection (RPC abort, timeout, and latency) by specifying the fault injection headers for E2E tests, and when requests go through microservices, faults will be applied according to the fault injection headers. The system features: (1) Safety: only injects failures on test tenancy requests, so production requests are safe; (2) Transparency: injects RPC errors on behalf of the destination service as if the destination service experiences actual failures; (3) Scalability: all microservices on Uber's Software Networking infrastructure are onboarded by default.

\begin{figure}[htbp]
\centering
\includegraphics[width=0.75\columnwidth]{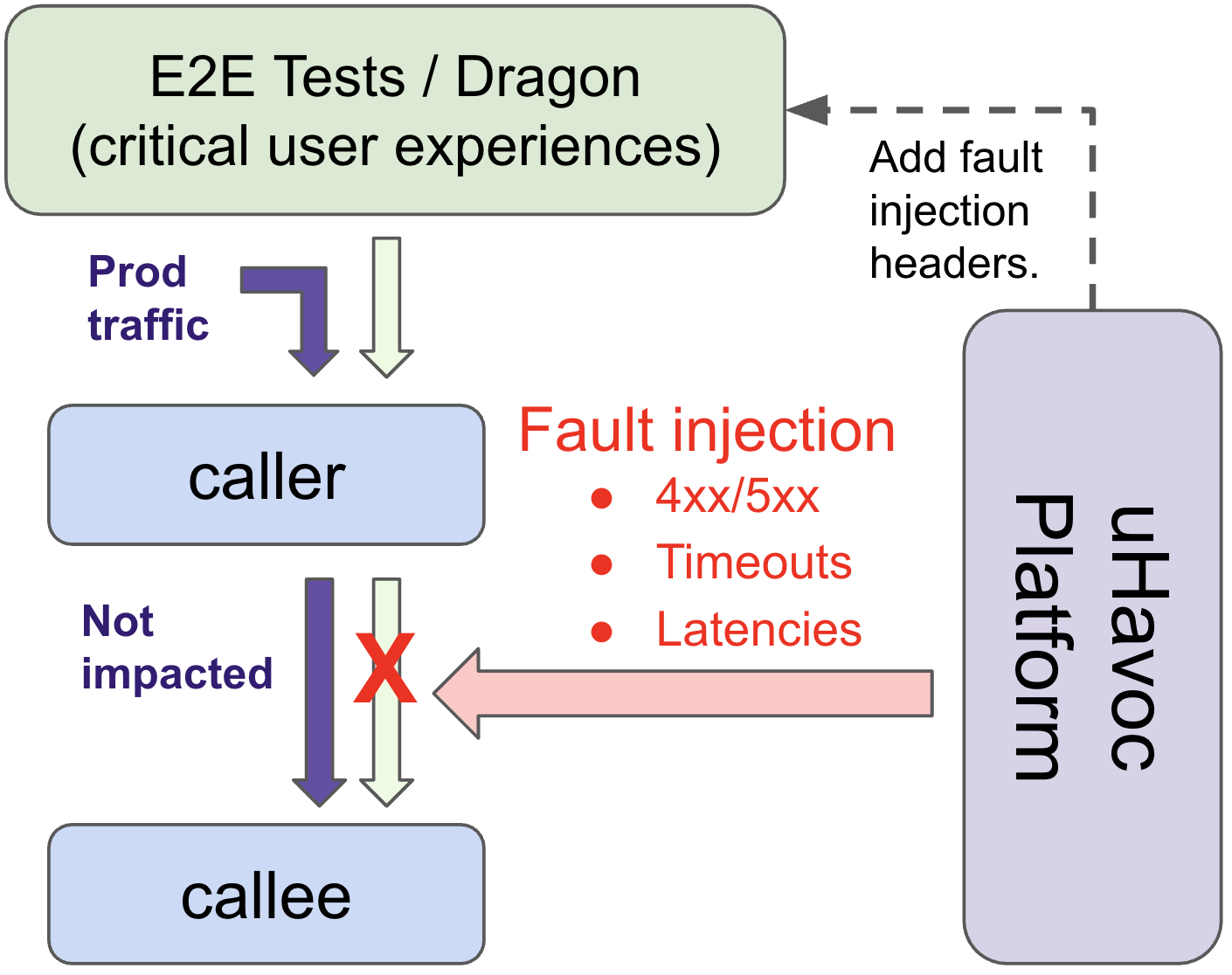}
\caption{uHavoc Platform Architecture showing fault injection process.}
\Description{Architecture diagram showing how uHavoc Platform injects faults between caller and callee services. Production traffic flows from E2E tests through the caller service, where the uHavoc Platform intercepts requests and injects faults including 4xx/5xx errors, timeouts, and latencies before reaching the callee service, while production traffic remains unimpacted.}
\label{fig:uhavoc}
\end{figure}

By Q1 2024, uHavoc had enabled the identification of 80+ resilience risks through various testing approaches, including Black-Box, BITs (Backend Integration Tests), Mobile, Web, and load testing. However, mobile testing remained limited by the need for manual test creation and execution.

\textbf{DragonCrawl:} DragonCrawl is an LLM-based system for resilient mobile end-to-end testing that uses textual representations of screens, goals, and previous actions to determine appropriate user actions, and then executes them~\cite{uber2024dragoncrawl}. As DragonCrawl executes actions, it collects screenshots, which are later analyzed to perform assertions. Assertions are formulated as visual question answering, where the inputs are the screen and a prompt, such as "Does the screen show drop-off locations?".

\begin{figure}[htbp]
\centering
\includegraphics[width=\columnwidth]{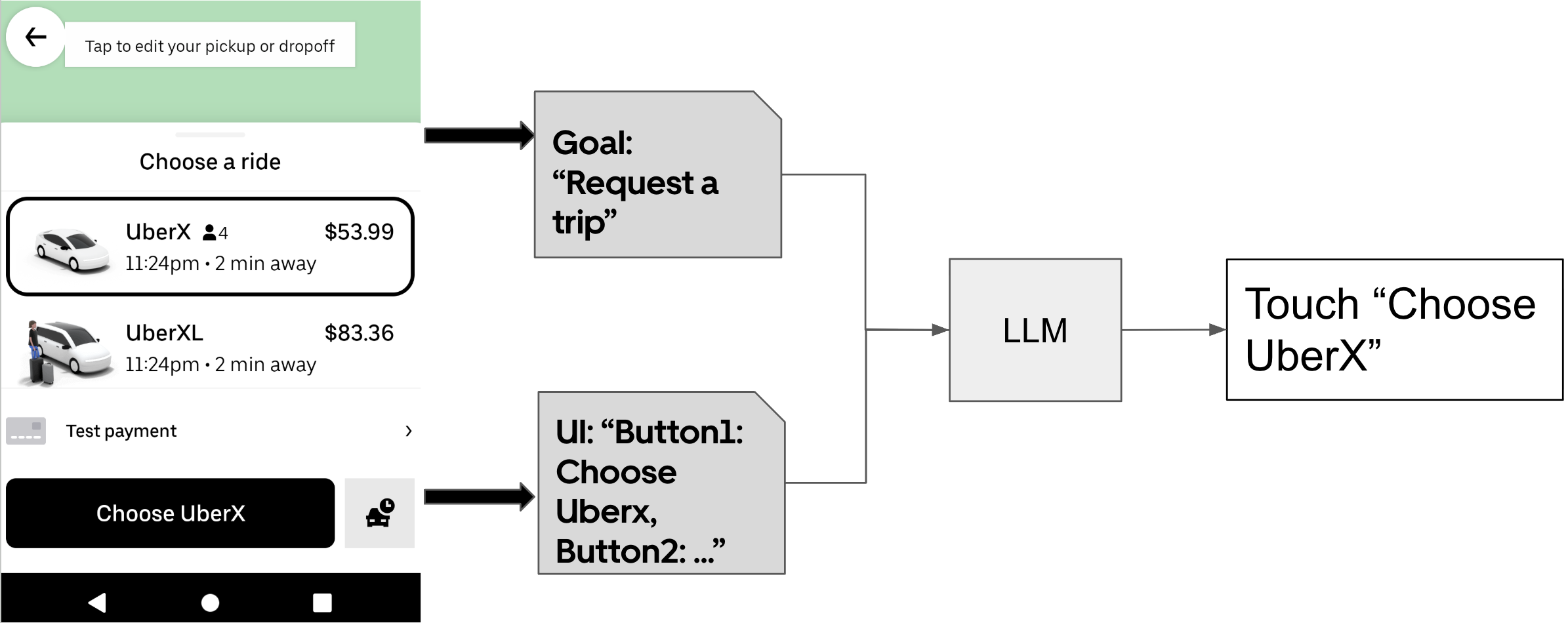}
\Description{Flow diagram of DragonCrawl showing how screens are parsed and content is analyzed by LLMs to determine actions.}
\caption{High-level overview of DragonCrawl.}
\label{fig:dragoncrawl}
\end{figure}

By analyzing what it sees on each screen and understanding the testing goal, DragonCrawl dynamically decides what actions to take rather than following pre-scripted sequences on a per test basis. This eliminates the need for engineers to hardcode separate test cases for every combination of language, location, fault-injection scenario, and UI variation - directly solving the combinatorial explosion that makes traditional testing approaches unscalable.

Unlike traditional testing methods that rely on brittle UI selectors or predetermined action sequences, DragonCrawl adapts to UI changes and flow variations in real-time, resulting in high test pass rates (99\%+) and low maintenance overhead. The system's resilience comes from its use of natural language understanding to interpret screen content and make contextual decisions, making it particularly suitable for testing in dynamic production environments where applications frequently change across different markets, languages, and failure conditions.

DragonCrawl's natural language understanding proves particularly valuable under chaos testing conditions. Traditional mobile testing requires explicit code to handle every unexpected delay, partially loaded UI element, or missing element ID, an approach that becomes intractable when fault injection introduces non-deterministic degradations. DragonCrawl interprets screens semantically: "Does this screen allow trip booking?" remains answerable when pricing shows "Calculating..." rather than a specific fare. This semantic understanding enables meaningful testing under the degraded conditions that chaos injection creates.

\section{System Design and Architecture}

Our system combines DragonCrawl's automated test execution capabilities with uHavoc's fault injection infrastructure. The architecture consists of three main components working in concert.

\subsection{Enhanced Test Case Setup}

Traditional DragonCrawl tests specify user flows using a simple checklist format that defines key screens to visit and actions to take. To enable chaos testing, we extended this specification to include fault injection parameters while preserving the existing test structure.

The enhanced test format builds upon DragonCrawl's existing components. The flow definition, end state assertions, and mid-state assertions remain unchanged from standard DragonCrawl tests. Flow definitions specify the sequence of user actions, end state assertions define criteria for successful test completion, and mid-state assertions verify that certain UI elements appear correctly throughout execution.

The key addition is a new fault configuration component that specifies which types of failures to inject, which service tiers to target, and the scope of fault injection. For example, a test might target all Tier 2+ services with timeout faults, or focus specifically on payment-related services with abort conditions.

This approach allows fine-grained control over failure scenarios while maintaining DragonCrawl's adaptive execution model. Rather than hardcoding specific fault conditions into individual test cases, the fault configuration acts as a parameter that can be varied across test runs. A single flow definition can be executed with different fault parameters to explore various failure scenarios, eliminating the need to write separate test cases for each combination of flow, city, and failure type.

The fault configuration integrates seamlessly with uHavoc's header-based injection mechanism. When a test specifies fault parameters, DragonCrawl automatically includes the appropriate headers in mobile application requests, ensuring that backend services receive and apply the intended fault conditions throughout the test execution.

\begin{figure}[htbp]
\centering
\includegraphics[width=\columnwidth]{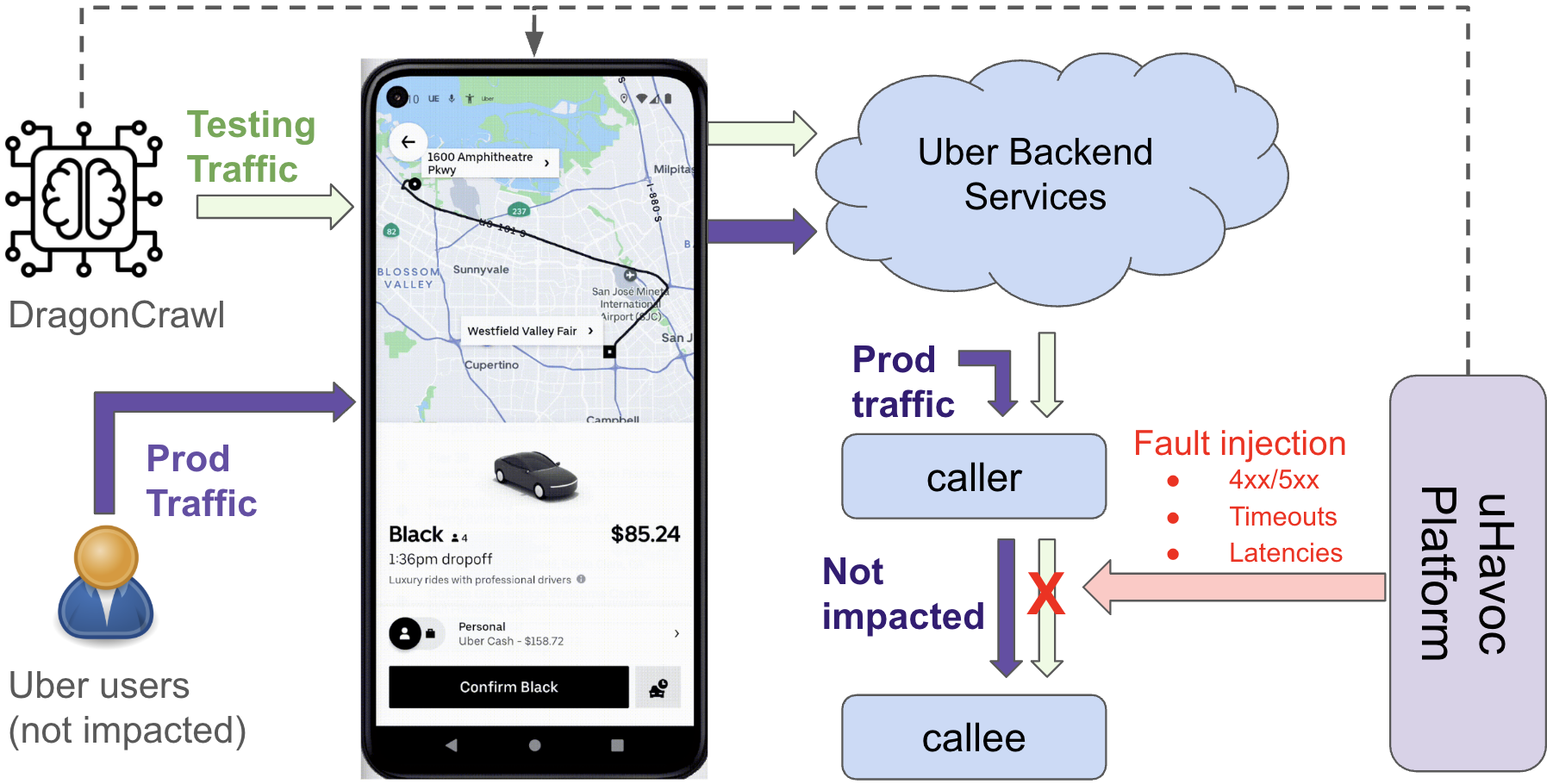}
\Description{Architecture diagram showing integration between DragonCrawl and uHavoc, with fault injection headers being passed from test execution through mobile apps to backend services.}
\caption{DragonCrawl and uHavoc integration architecture.}
\label{fig:integration1}
\end{figure}

For example, the core trip flow test now includes:
\begin{itemize}
\item \textbf{Flow Definition:} Test steps (i.e. login, accept the terms and conditions, choose a dropoff location, etc.)
\item \textbf{End state assertions:} Asserts whether the current screen matches a criteria to end the crawling phase and move to the assessment step (e.g., "Does the screen show both tipping options for the driver and a star-based rating prompt for the trip")
\item \textbf{Mid state assertions:} Asserting whether certain elements appear across the entire run (i.e. discounts, driver selfie, vehicle make and model, etc.). This is made against a mosaic of all app screens stitched together
\item \textbf{Fault Configuration:} Headers specifying fault types, target service tiers, and injection scope
\end{itemize}

The fault configuration allows fine-grained control over which services are targeted and what types of failures are injected, enabling both broad resilience validation and focused testing of specific dependency relationships.

\subsection{Automated Root Cause Analysis and Ticketing System}

Prior to creating our mobile chaos testing platform, resilience risks would be triaged manually which would not scale. Diagnosing the offending service and/or RPC call used to require the full attention of senior engineers with long tenure at Uber (4+ years) with enough context of Uber's complex microservice graph, which limited the pool of triagers to less than 10 or 15 engineers at the company. Thus, in order to run this at scale, and to future-proof Uber in case of attrition, we decided to create a system to automatically root cause resilience risks.

\textbf{Mobile Application Instrumentation Requirements:}
When a mobile havoc test fails, comprehensive instrumentation data is critical to understand the root cause and raise tickets for service owners to address resilience risks:

\textbf{Mobile Network Log Integration:} The in-app bug reporter within the mobile app is triggered after every DragonCrawl test failure. The report contains a buffer of each network request and response made by the app.

\textbf{Screen Transition Integration:} The triaging engineer needs to replay the execution flow to visualize each screen shown to the model, the action taken by the model, and the reasoning provided for that specific action.

\textbf{Jaeger Tracing Integration:} Enhanced mobile applications to enable distributed tracing during chaos tests, allowing for precise attribution of failures to specific backend services.

\textbf{Test Account Configuration:} Modified test account setup to enable header injection capabilities and appropriate experiment flags for chaos testing scenarios.

\begin{figure}[htbp]
\centering
\includegraphics[width=0.85\columnwidth]{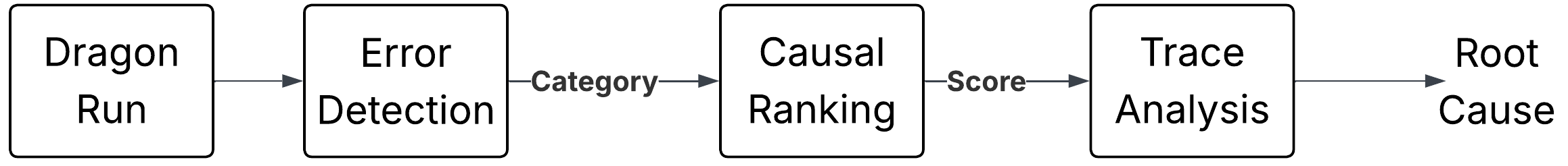}
\Description{Pipeline diagram showing three phases of automated root cause analysis: error detection, causal ranking with statistical analysis, and trace analysis leading to root cause identification.}

\caption{Automated Root Cause Analysis Pipeline.}
\label{fig:rca}
\end{figure}

\textbf{Root Cause Analysis Heuristic:}
Our approach combines traditional statistical analysis with modern large language model (LLM) capabilities to perform automated root cause analysis in mobile application testing. The system operates in three primary phases: error detection, causal ranking, and trace analysis.

The RCA service first attempts to detect the presence of error messages or missing UI components in each screen across multiple apps. The error detection occurs on the XML screen hierarchy output by the mobile app driver. Each screen is classified using a hybrid approach that first applies regex pattern matching to identify common error phrases, followed by LLM-based classification for more nuanced error states.

The causal ranking phase employs a multi-factor scoring model that integrates frequentist statistical analysis with semantic analysis. Network requests associated with error screens are analyzed through an LLM-based relevance categorization system that classifies each API endpoint into four categories with corresponding weights: direct (3.0), indirect (2.0), supporting (1.2), and unrelated (0.3).

The core ranking heuristic combines multiple factors into a single causal score using the formula: \textbf{Score(causal) = F(status\_code) × (1 - normalFailureRate) × F(tier) × F(category)}, where F(status\_code) assigns 1.0 to server errors (5xx), 0.5 to client errors (4xx), and 0.2 to success responses (2xx); normalFailureRate represents the frequency with which a specific request pattern appears in successful baseline runs; F(tier) weights services by criticality (1.0 for tier-0, 0.9 for tier-1, 0.7 for tier-2, 0.4 for tier-3, and 0.1 for tier-4/5); and F(category) applies the LLM-derived semantic weights. These parameter values were established through empirical observation and have proven consistent across daily execution of all 47 core flows.

\section{Implementation Challenges and Solutions}

During deployment, we encountered several technical challenges that required iterative solutions. One of the most significant issues was distinguishing between genuine resilience problems and environmental noise. Mobile testing environments are inherently dynamic---live changes occur constantly across mobile apps, backend services, and platform infrastructure. These changes could trigger test failures that had nothing to do with our injected faults, creating false positives that would undermine confidence in the system.

To address this, we implemented baseline comparison testing. For every chaos test run, we execute an identical control test without fault injection. This approach allows us to isolate failures caused by actual service degradation from those caused by unrelated system changes.

As we scaled up testing, our root cause analysis pipeline began overwhelming Uber's Jaeger tracing infrastructure. Processing an average of 1,000 DragonCrawl runs daily meant sampling over 200,000 traces per day through Jaeger's public FindTraces endpoint. The downstream systems couldn't handle this load, leading to out-of-memory errors and rate limiting that blocked our analysis pipeline.

Rather than requesting additional Jaeger capacity, we pivoted to leveraging mobile devices' built-in network logging capabilities. We expanded the existing in-app bug reporter to capture network request/response data directly from test devices, reducing our dependence on centralized tracing infrastructure while actually improving data fidelity.

The AI-driven nature of DragonCrawl introduced complexities. The LLM entered problematic loops ~2\% of executions, either believing it had not completed a step or repeatedly taking ineffective actions. We implemented cycle detection to prompt the model toward alternative approaches, eliminating these loops. However, the model still occasionally takes ineffective actions before finding the correct path, manifesting as increased latencies and failures. These failures account for 0.4\%, and because they are rare and easily distinguishable from genuine resilience risks, we do not triage them manually. Thus, AI complexities impose essentially zero maintenance overhead while achieving 99.3\% pass rate and eliminating the combinatorial explosion requiring 20,000+ manually-written test cases.

Finally, some of Uber's most critical flows span multiple applications simultaneously. The core trip flow, for example, involves a rider using one app instance while a driver uses another to accept and complete the ride. When failures occurred in these multi-app scenarios, our root cause analysis needed to aggregate network activity across both applications to identify which backend error was most likely responsible for the overall test failure.

\section{Evaluation}

In this section, we evaluate our system's performance prior to launch. DragonCrawl initially used MPNet~\cite{uber2024dragoncrawl}, but pass rates declined as flow complexity and context window requirements increased. All LLM-based components (action selection, visual question answering, and root cause analysis) now use GPT-4o, selected for its superior performance on screen understanding tasks. We measured DragonCrawl's performance with and without fault injection across model accuracy, pass rate, and latency, then assess our root cause analysis pipeline.

\subsection{Model Performance}

In our blog~\cite{uber2024dragoncrawl}, we discuss in detail how we framed the performance metrics of DragonCrawl as retrieval tasks, similar to metrics used in recommendation systems. Since the goal is for DragonCrawl to suggest the right UI action to take (e.g., which button to tap, which field to fill) ideally in its first attempt, the right metrics to assess DragonCrawl for offline evaluation is precision@k, with k being a small integer, ideally 1.

\begin{table}[htbp]
\centering
\small
\setlength{\tabcolsep}{4pt}
\caption{DragonCrawl precision@k performance across fault injection scenarios}
\label{tab:dragon_precision}
\begin{tabular}{lccccc}
\toprule
\textbf{Metric} & \textbf{Baseline} & \textbf{T5} & \textbf{T4} & \textbf{T3} & \textbf{T2} \\
\midrule
Precision@1 & 0.9723 & 0.9716 & 0.9716 & 0.9512 & 0.9408 \\
Precision@2 & 0.9623 & 0.9641 & 0.9641 & 0.9473 & 0.9408 \\
Precision@3 & 0.9423 & 0.9398 & 0.9398 & 0.9318 & 0.9408 \\
\bottomrule
\end{tabular}
\end{table}

\autoref{tab:dragon_precision} summarizes our observations on precision@k for k=1,2, and 3 for the baseline and executions with T5-T2 failures (where T5 represents lowest-priority services and T2 represents non-critical but important services, as defined in Section 2.3) for the trip request flow, where a rider controlled by a DragonCrawl agent requests a trip and a driver controlled by a DragonCrawl agent accepts and completes the trip, which we executed 300 times, spanning over 6000 actions on UI elements for each fault injection scenario. Our observations for this particular were inline with what we expected: When injecting faults to the services with lowest priorities (T5 and T4), there were no noticeable changes to precision@k for k=1,2,3. However, once failures were injected to T3 services, we started to see performance decrease slightly, given that some UI elements started to take longer to load or respond. Similarly, when injecting backend faults to T2 failures, performance degraded again because delays would be even longer. For T3 and T2 cases, DragonCrawl would wait, but if the delays were too long it would try to find other routes to request the trip, which resulted in negative scoring in precision@k. This non-deterministic behavior in itself would be problematic for regular regression testing, but it is exactly what chaos testing requires.

\subsection{Latency}

\begin{table}[htbp]
\centering
\small
\setlength{\tabcolsep}{4pt}
\caption{DragonCrawl latency performance across fault injection scenarios}
\label{tab:dragon_latency}
\begin{tabular}{lccccc}
\toprule
\textbf{Metric} & \textbf{Baseline} & \textbf{T5} & \textbf{T4} & \textbf{T3} & \textbf{T2} \\
\midrule
P50 Latency (s) & 166.3 & 168.1 & 164.0 & 179.8 & 188.2 \\
P95 Latency (s) & 204.5 & 203.9 & 206.6 & 223.2 & 240.8 \\
P99 Latency (s) & 212.7 & 213.1 & 213.8 & 240.1 & 260.2 \\
\bottomrule
\end{tabular}
\end{table}

\autoref{tab:dragon_latency} summarizes how latency changes with fault injection. This metric also follows the pattern observed in model performance: Minimal changes in latency for the baseline and for fault injection in services of the lowest priority. However, additional delays do come up as we inject backend faults to T3 and T2 services. For the P50 latency, injection of backend faults results in approximately 13 extra seconds relative to the baseline because some screens or UI elements take slightly longer to load or respond, for T2 failures, the P50 latency is almost 20 seconds longer, due to longer delays. The P95 and P99 show a wider gap between the baseline and the T3 and T2 fault injection scenarios. The reason for that is that P95 includes the cases where DragonCrawl encountered pop-ups, offers, and other extra screens that were not seen in the P50, and because there are backend faults, the UI elements in these screens take longer to respond, so the delays end up cascading.

\subsection{Pass Rate}

\begin{table}[htbp]
\centering
\small
\setlength{\tabcolsep}{4pt}
\caption{DragonCrawl pass rate across fault injection scenarios}
\label{tab:dragon_passrate}
\begin{tabular}{lccccc}
\toprule
\textbf{Metric} & \textbf{Baseline} & \textbf{T5} & \textbf{T4} & \textbf{T3} & \textbf{T2} \\
\midrule
Pass rate & 0.993 & 0.993 & 0.993 & 0.993 & 0.99 \\
\bottomrule
\end{tabular}
\end{table}

Finally, we computed the pass rate for the scenarios previously described. A test "passes" when DragonCrawl successfully reaches the defined end state (e.g., the trip rating screen) within the timeout window, indicating the critical user flow remained functional despite injected faults. The results, which can be seen in \autoref{tab:dragon_passrate}, follow the trend we observed for the offline model performance metrics: Pass rate does not change when backend faults are injected to T5-T3 failures, but they drop slightly for T2 failures. Thus, in spite of the unexpected delays and unresponsive UI observed in the prior discussion, DragonCrawl was able to achieve 99\% or higher pass rate.

\subsection{Visual Question Answering (Assertion) Performance}

To validate that our assertion system correctly identifies whether UI elements are present or absent, we formulated it as a binary classification problem. We collected 6000 screens and evaluated two assertions per screen: one assertion that should return true and one that should return false. Assertions are framed as questions regarding whether or not something appears on the screen with a true/false answer. We evaluated both LlaVa-NeXT and GPT-4o for VQA tasks, ultimately selecting GPT-4o for production deployment due to its superior performance on true/false binary classification.

\begin{table}[htbp]
\centering
\small
\setlength{\tabcolsep}{4pt}
\caption{Visual Question Answering confusion matrix}
\label{tab:vqa_confusion}
\begin{tabular}{lcc}
\toprule
 & \textbf{Predicted Positive} & \textbf{Predicted Negative} \\
\midrule
\textbf{Actual Positive} & 0.99967 & 0.00033 \\
\textbf{Actual Negative} & 0 & 1.0 \\
\bottomrule
\end{tabular}
\end{table}

The results from these experiments were fairly positive and were not surprising (see \autoref{tab:vqa_confusion}). Since 2018, character and text recognition models have achieved $\sim$99\% accuracy, even in low resolution images~\cite{ocr_accuracy_ref1,ocr_accuracy_ref2}.

Something worth noting is that our decision to frame assertions as true/false questions is grounded in the fact that visual question answering that require more complex reasoning has much lower accuracy, with GPT-4o achieving 77.57\% accuracy on the Complex Reasoning Visual Questioning Answering dataset~\cite{corevqa}.

\textbf{Validating Test Fidelity Under Chaos:} A critical concern with adaptive AI testing is whether DragonCrawl masks real resilience issues by finding unintended workarounds. Two mechanisms prevent this: First, tests have explicit goals, such as completing a trip, ordering food, receiving driver confirmation. If these screens are not reached, the test fails. When backend failures block core functionality, DragonCrawl cannot work around them; the "golden screen" simply never appears. This goal-based validation is reliable regardless of which path the test takes.

Second, visual assertions validate critical UI elements at key screens: trip price, driver information, payment confirmation. We deliberately do not assert every element. Promotional banners and feature suggestions would be nice to validate, but they have minimal business impact and take months to fix. At Uber's scale, asserting everything creates noise that obscures critical failures. Our focused approach detects resilience issues that matter while avoiding alert fatigue from cosmetic problems. The 23 issues we found—12 blocking trips/orders, 2 causing crashes—show this approach captures genuine failures rather than masking them.

\subsection{Root Cause Analysis Accuracy}

To evaluate the performance of Root Cause Analysis, we framed it as a ranking and recommendation problem, similar to offline metrics used in DragonCrawl. The reason for that is that the output of the root cause analysis system is a ranked list of API/RPC calls, ranked by causality. Ideally, the first RPC call recommended by the system should be the one causing the resilience risk, and the lower the offending RPC call is in the ranked list, the longer a developer takes to root cause. Thus, we chose precision@k to measure performance.

\begin{table}[htbp]
\centering
\small
\setlength{\tabcolsep}{4pt}
\caption{Root Cause Analysis precision@k performance}
\label{tab:rca_precision}
\begin{tabular}{ccc}
\toprule
\textbf{Precision@1} & \textbf{Precision@3} & \textbf{Precision@5} \\
\midrule
0.50 & 0.71 & 0.88 \\
\bottomrule
\end{tabular}
\end{table}

\autoref{tab:rca_precision} shows the results of our evaluation. In short, what our experiments indicate is that while the top RPC call may not be the root cause, users may not need to search too far to get to the root cause.

Since every root cause analysis execution takes approximately 2 minutes, needing to run the tool multiple times would result in customers getting to the root cause in approximately 10 minutes, which would be a substantial improvement over the approximately 3 hours we witnessed engineers spend root causing.

We debated whether or not to invest more resources into increasing precision@1, but given that we did not expect the volume of resilience risks per month to be large, we decided not to. As mentioned above, it only takes retries to get to the root cause.

\section{Operational Experience and Results}

Since Q1 2024, our integrated DragonCrawl + uHavoc chaos testing system has been operational in production, running nightly automated experiments across Uber's core mobile flows. Across 47 critical flows, the system has consistently achieved a 99.27\% pass rate, with observed latencies matching those measured in pre-deployment evaluation (P50 $\approx$166--188s; P95 $\approx$200--240s). These results confirm that the platform is both stable and safe for continuous operation while providing resilience validation at scale.

\textbf{Testing Coverage:}
\begin{itemize}
  \item \textbf{Flow Coverage:} Tests execute across all 47 DragonCrawl-supported core flows spanning Rider, Driver, and Eats applications
  \item \textbf{Fault Scenarios:} Initial focus on Tier 2+ (non-critical) service failures and timeouts, with plans for expansion to more granular failure types
  \item \textbf{Execution Frequency:} Nightly execution schedule with capability for ad-hoc testing as needed
\end{itemize}

\subsection{Issue Discovery and Impact}

Since deployment, our automated mobile chaos testing has identified over 23 resilience risks (3 open, 3 in progress, and 17 resolved) across Uber's mobile applications and backend infrastructure. The distribution of issues provides valuable insights into common resilience anti-patterns:

\textbf{Dependency Violations (70\% of issues):} The majority of identified problems were cases where Tier 2+ (non-critical) service failures caused degradation or failure of critical user flows. These violations conflict with Uber's architectural principles that higher-tier service failures should not impact core business functionality.

\textbf{Timeout Misconfigurations (15\% of issues):} Several cases where service timeout configurations were inappropriate for mobile user experiences, leading to poor performance during degraded service conditions.

\textbf{Fallback Implementation Gaps (10\% of issues):} Instances where mobile applications lacked appropriate fallback mechanisms for specific service failure scenarios, resulting in poor user experiences.

\textbf{Circuit Breaker Issues (5\% of issues):} Cases where circuit breaker configurations were inappropriate for mobile access patterns, leading to either too-aggressive failure isolation or insufficient protection.

\begin{figure}[htbp]
\centering
\includegraphics[width=0.85\columnwidth]{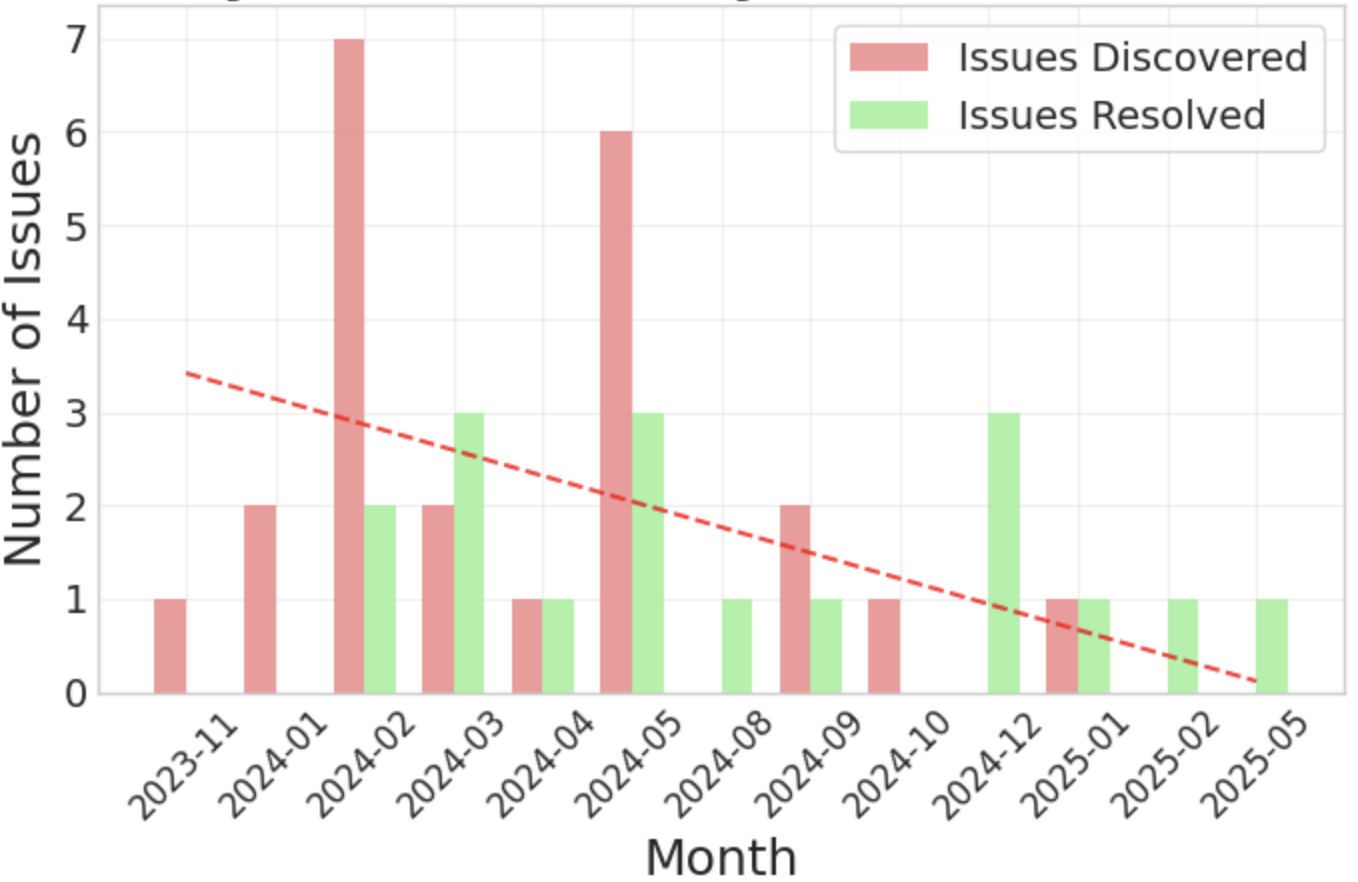}
\Description{Line graph showing monthly resilience issue discovery and resolution from November 2023 to May 2025, with peak discovery in early 2024 and declining trend thereafter.}
\caption{Monthly resilience issue discovery and resolution timeline.}
\label{fig:monthly}
\end{figure}

Figure~\ref{fig:monthly} shows the number of resilience issues discovered and resolved every month between November 2023 and May 2025. Initially, we encountered the cold start problem: there was skepticism in our solution, and it was difficult to find owners for the issues we discovered. However, starting in February 2024, when we started to communicate our work more broadly and gained believers, teams started to resolve the issues we would discover. As expected, the majority of the issues were discovered between February 2024 and May 2024, but we continued executing the tests on a nightly basis to detect any regressions/newly introduced resilience risks. As the the dotted trendline shows, because we already discovered many resilience risks, we expect to only discover regressions in resilience moving forward.

A pattern that we noticed was in issue resolution is that teams with more than one issue assigned would resolve tickets faster. They would create sprints completely dedicated to operational excellence, during which they would close 2-4 resilience risks.

\begin{figure}[htbp]
\centering
\includegraphics[width=0.85\columnwidth]{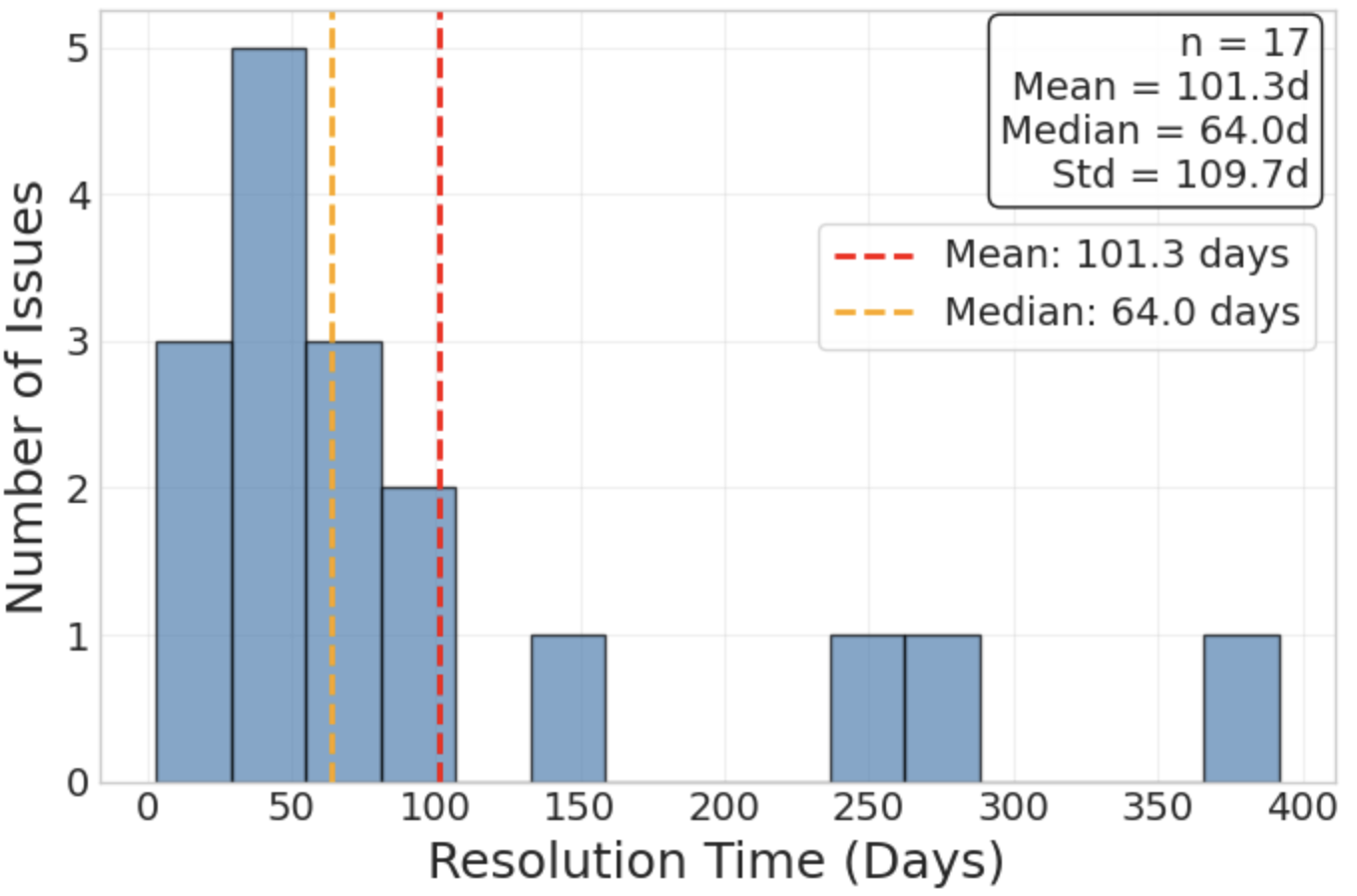}
\Description{Histogram showing distribution of issue resolution times in days, with median at 64 days and most issues resolved within 45 days since February 2024.}
\caption{Distribution of issue resolution times.}
\label{fig:resolution}
\end{figure}

Regarding resolution times, we observed that it is much longer than typical backend or mobile bugs. Since these discoveries do not represent issues affecting customers at the time of discovery, teams may postpone fixing these issues to deliver on their quarterly commitments. Furthermore, as we mentioned earlier, to determine the appropriate and durable fix, the engineer owning the resilience risk may need to review 10s or 100s of RPC calls, which requires understanding microservices they do not own. Thus, the time required to resolve these issues is much longer than that of typical backend or mobile bugs, whose SLA is 1 day for bugs that would prevent customers from requesting trips/meals. The median resolution time is 64 days, with issues discovered since February 2024 typically resolved in under 45 days.

\begin{figure}[htbp]
\centering
\includegraphics[width=0.85\columnwidth]{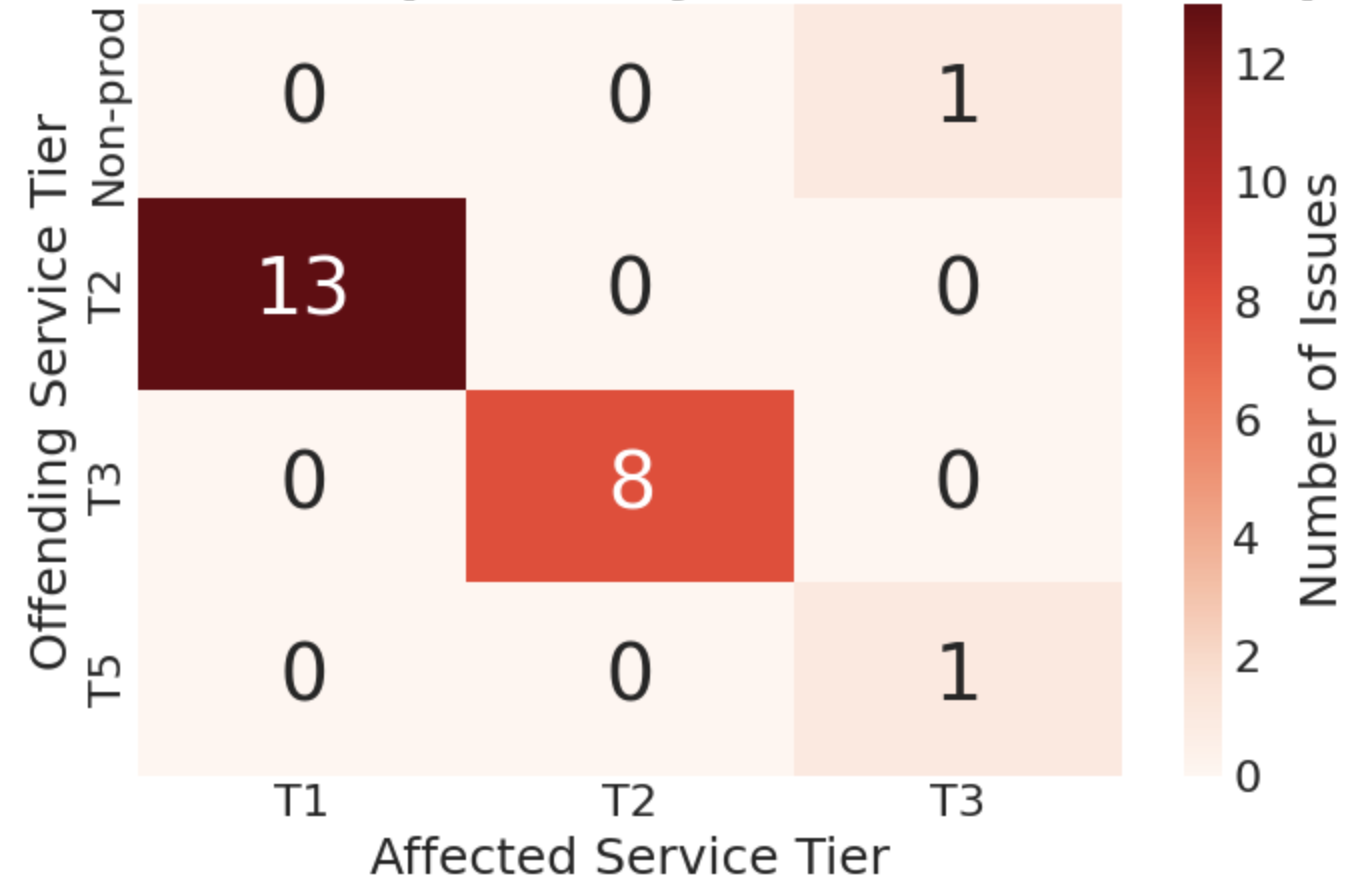}
\Description{Heatmap showing service tier interactions for resilience risks, with hottest spots at tier-2 services causing tier-1 failures and tier-3 services impacting critical flows.}
\caption{Service tier interaction heatmap showing resilience risks.}
\label{fig:heatmap}
\end{figure}

We only tested in scenarios where fault injection should not have caused the user flow under test to fail, and Figure~\ref{fig:heatmap} illustrates the priority of the services causing resilience risks. The hottest spot in the heatmap is the case where tier 2 (non-critical) services caused failures is tier1 (critical) services, which occurred for 13/23 resilience risks. The second hottest spot is in the middle of the heatmap, where tier 3 services, which should not have any customer impact, would actually cause user flows such as requesting a trip to fail. Finally, we had one case where a failure in a tier 5 service (lowest priority possible) took down a critical payments feature at Uber in our tests. Similarly, we discovered a case where a service in a staging environment was communicating with a production service that would also impact payments.

\begin{table}[htbp]
\centering
\small
\setlength{\tabcolsep}{4pt}
\caption{Issue distribution by severity level}
\label{tab:severity_distribution}
\begin{tabular}{ccc}
\toprule
\textbf{Highest Severity} & \textbf{High Severity} & \textbf{Moderate Severity} \\
\midrule
12 & 9 & 2 \\
\bottomrule
\end{tabular}
\end{table}

Finally, most of the issues that we detected were high severity issues. There were 12/23 issues that would prevent users from requesting trips or meals or drivers from receiving and completing trips. These kinds of issues are considered the most severe issues at Uber. There were 9/12 issues where payments or Uber memberships would be affected, which could indirectly affect trips and meals, and thus are of high severity. There were 2/23 issues that would affect Uber products with much lower revenue than trips and meals, but still produce customer impact and therefore are also considered severe.

Although uHavoc can detect resilience risks by working with other end-to-end testing systems, these 23 issues that we detected were not detected through other means. Furthermore, 2/23 issues that we detected would cause the Uber apps to crash, which means that they could only be detected through mobile chaos testing.

\subsection{Operational Benefits}

\textbf{Reduced Manual Testing Overhead:} The automated system eliminates the need for manual chaos testing coordination, reducing engineering overhead while increasing frequency of testing in 47/61 critical user flows, with the exception of flows that require validating legal documents and user pictures (14/61 flows), which are executed by manual testers.

To manually execute a typical fault injection test, manual QA engineers would need to manually create at least one test account and configure it so that it contains the right experiment flags and fault injection, which takes approximately 4 minutes per account. On the other hand, our chaos testing platform automates account creation and fault-injection setup so that this entire provisioning process can be done with 2 RPC calls, with total P99 latency under 7 seconds.

This automation enables testing at unprecedented scale. Our system has executed over 180,000 automated chaos test runs since deployment. Manual execution of our test suite would require significant human effort per test:
\begin{itemize}
    \item Account setup: 4 minutes
    \item Fault injection configuration: 2 minutes  
    \item Supervised test execution: 7 minutes
    \item Total per test: 13 minutes
\end{itemize}
Across 180,000 tests, this represents 39,000 hours of automated manual work that has been automated away.

Mobile chaos testing started as an expensive, occasional activity limited by human resource constraints, but with our work it has become a routine nightly validation process. The 47 automated flows now execute nightly via DragonCrawl, while the remaining 14 flows requiring manual validation continue to be tested weekly by human QA engineers. In the past, mobile resilience testing occurred on an ad-hoc basis due to manual coordination overhead, but automated execution now provides frequent validation of architectural resilience principles across the majority of our critical user flows.

\textbf{Easy Adoption:} The Uber Failover Architecture team utilizes DragonCrawl and uHavoc to simulate failover drills. This team creates a list of non-critical services to be dynamically terminated during a region failover to make room for critical trip and order serving services, and prior to those exercises, they trigger the DragonCrawl flows with faults injected to the target list to simulate what would happen during real drills. This process has been executed bi-weekly for 27 drills between July 2024 and September 2025, and the pass rate of our tests has been 99.27\%.

\textbf{Faster Issue Detection}: Automated execution and analysis enables identification of resilience issues during the development cycle rather than during production incidents.

\textbf{Improved Attribution:} Integration with distributed tracing provides precise attribution of mobile issues to specific backend services, accelerating resolution and reducing debugging time. Since launching the root cause analysis tool, non-experts report being able to root cause resilience risks in under 5 minutes, which is a task that require multiple hours from a tenured engineer with high context.

\textbf{User Experience Visualization}: In contrast to conventional unit tests and integration tests, DragonCrawl enabled uHavoc tests not only reveal potential risks in service resilience, but also offer visual representations of the customer experience through failed mobile screenshots and transitions. This allows for a better understanding of how the injected failures would impact production.

\subsection{Lessons Learned}

\textbf{Start Simple to Build Trust.} Beginning with well-understood failure scenarios helped gain organizational confidence. Expanding too quickly would have risked credibility if early tests produced noisy or ambiguous results.

\textbf{Safety is Non-Negotiable.} Fault injection must operate under strict tenancy and isolation controls. We learned that clear guardrails (e.g., test-only traffic separation) were essential to organizational adoption, as engineers were initially skeptical about any chaos experiments running against production environments.

\textbf{Integration Requires Significant Engineering.} Even though DragonCrawl and uHavoc were mature systems, combining them required non-trivial engineering effort. Timing coordination, account provisioning, and cross-system communication had to be carefully redesigned to avoid false positives.

\textbf{Invest in Observability Beforehand.} Our automated root cause analysis depended on rich tracing (Jaeger), network logs, and screen transitions. Incomplete tracing or inconsistent service tier tagging would have produced misleading results. Organizations attempting similar work should prioritize distributed tracing and proper service tier classification first.

\textbf{Balance Scope with Practicality.} While our system validated resilience across 47 critical flows, certain flow variations (e.g., airport-specific experiences) remained out of scope. A key lesson was to identify \emph{representative} flows that capture the majority of risk, rather than aiming for exhaustive coverage.

\textbf{Third-Party and Infrastructure Gaps.} Because uHavoc operates at the RPC layer, infrastructure-level degradations (e.g., packet loss) and third-party service failures (payments, maps) remained untested. This was a conscious trade-off to ensure safety. Others should carefully scope what types of failures can and cannot be safely injected.

\textbf{Organizational Dynamics Matter.} The technical system surfaced issues, but resolution required cross-team ownership. We observed that teams closed issues faster when multiple were assigned, often dedicating a sprint to resilience. Building explicit processes for triage and ownership proved just as important as the technology itself.

\section{Related Work}

Chaos engineering became popular in the industry with Netflix's contributions. Chaos Monkey terminates production instances to validate service resilience, while Failure Injection Testing (FIT) precisely injects request-level faults to selected components and user cohorts in production experiments~\cite{netflix_chaos_monkey,netflix_fit}. These efforts established the current pattern of hypothesis-driven failure injection with strict blast-radius controls.

Subsequent platforms generalized fault injection beyond a single company. Gremlin productionized chaos experiments (latency, aborts, resource pressure) and advocated integrating them into reliability programs and CI/CD pipelines~\cite{gremlin}. Google has documented extensive chaos testing for Spanner in production-like environments to validate database availability under realistic failures~\cite{google_spanner_chaos}.

Closer to the service graph, Service-level Fault Injection Testing (SFIT) explores failing specific RPCs to expose resilience bugs early. DoorDash reported applying SFIT to microservice endpoints, building on the Filibuster research which combines static/concolic analysis with dynamic reduction to surface failure interactions efficiently~\cite{doordash_filibuster,filibuster_paper}. Lyft's Chaos Experimentation Framework (CEF), built atop Envoy/Clutch, makes fault injection self-service and ties it to deployments. Notably, Lyft highlights ensuring \emph{client resiliency}, i.e., exercising mobile client endpoints under backend degradation to verify fallbacks—an explicit nod to the mobile UX impact, though the experiments are orchestrated at the mesh/service layer~\cite{lyft_cef}.

In parallel, mobile testing research has advanced significantly in automation and maintenance reduction. Model-based testing, automated test generation, and vision-language models have improved the robustness of UI automation, but these efforts have focused on functional correctness rather than resilience under backend failures~\cite{mobile_testing_survey,model_based_mobile,automated_mobile_generation,vision_mobile_testing}.

Academic and practitioner surveys synthesize these practices, but they remain focused on server-side failure models rather than end-to-end mobile experience validation with fault injection~\cite{rosenthal_book,error_models}. 

Prior art predominantly injects faults in backend services and occasionally targets mobile endpoints at the API layer. We did not find publications of a system that (1) couples automated, LLM-driven mobile UI test execution (resilient to UI changes) with (2) systematic service-level fault injection and (3) automated RCA that attributes mobile failures to specific backend RPC calls, and does so continuously across tens of thousands of nightly runs for dozens of core flows. Our work fills that gap by making the mobile experience the primary observation point while retaining the rigor and safety of service-layer chaos experiments. 

\section{Limitations}

Despite the successful implementation of DragonCrawl and uHavoc, which allowed for safe and automated mobile testing on a large scale, we encountered several main challenges with the system.

\textbf{The Trade-off for Safe Chaos Testing:} To ensure the safety of chaos testing, uHavoc strictly operates on testing tenancy requests, so that production requests are separated. This means fault injection can only happen at the service request level. The system will not be able to perform infrastructure-level chaos testing to verify zonal packet loss network degradation. This is a trade-off that we made consciously to ensure Uber's customers are well protected from any forms of chaos testing.

\textbf{Limited Scope for Services on the Service Mesh:} Since the uHavoc platform operates on the service level, it can only apply faults (aborts, timeout, latency) between two RPC microservices. This makes it difficult to verify failure resilience for storage related services (e.g. Kafka, Flink, or Cassandra), or services that are not on Uber's service mesh infrastructure. Further, the system cannot inject faults into third-party services (payment processors, mapping services, external APIs) that Uber depends on but does not control. This creates coverage gaps for testing resilience to external service failures that commonly impact mobile applications in production.

\textbf{Coverage of Flow Variations:} Even though DragonCrawl scalably covers most of the critical user flows (47 total), it is hard for it to cover different variations of a single flow. One example is location based airport experiences. DragonCrawl covers this flow, but it is not scalable to execute the same Dragon flow for each airport where Uber operates, since there are thousands of airports in the US alone. As a result, if an airport flow has its unique resilience risks, DragonCrawl and uHavoc will not be able to cover it if the test is not configured to execute for that city.

\textbf{Proper Tagging of Service Tiers:} A portion of our success comes from the rigorous effort teams have put in to correctly identify the priority of their services. Without this effort, our platform would have created much more resilience risks and teams would have needed to spend time investigating RPC calls and Jaeger traces, when in reality the failures would be caused simply by improper tagging. Organizations trying to replicate our work should spend some time first properly tagging their services.

\textbf{Organizational Maturity Requirements:} Beyond service tier tagging, successful implementation requires mature DevOps practices including comprehensive monitoring, incident response procedures, and a culture that embraces controlled failure testing. Organizations lacking these foundational elements may struggle to effectively utilize the insights generated by the system.

\textbf{Distributed Tracing Dependencies:} The system's failure attribution capabilities rely heavily on comprehensive distributed tracing (Jaeger) being enabled across all services. Services that lack proper tracing instrumentation or have incomplete trace propagation may result in attribution gaps, making it difficult to correlate mobile symptoms with specific backend failures. This creates blind spots in the analysis and may lead to unresolved issues.

\section{Conclusion}

Our experience integrating DragonCrawl with uHavoc demonstrates that automated mobile chaos testing can operate effectively at production scale. The system has executed over 180,000 tests since Q1 2024, a volume that would require approximately 39,000 hours of manual effort, a scale unachievable through traditional testing approaches.

The practical results validate the approach. We identified 23 resilience risks, predominantly architectural dependency violations where failures in non-critical services degraded core user flows. Twelve of these issues were severe enough to prevent trip requests or food orders, and two caused application crashes that backend testing alone could not detect. Automated root cause analysis reduced debugging time from hours to minutes, enabling engineers without deep system knowledge to diagnose resilience issues independently.

Several factors proved critical to success. Safety mechanisms with strict traffic isolation were non-negotiable for organizational adoption. Starting with well-understood failure scenarios built confidence before expanding scope. Automated attribution provided actionable insights rather than vague alerts. However, replicating this approach requires substantial infrastructure prerequisites: comprehensive distributed tracing, accurate service tier classification, and organizational willingness to address discovered issues.

The transition from manual, ad-hoc mobile resilience testing to continuous automated validation represents a significant operational shift. Our work shows this transformation is achievable at scale, though not without cost. As mobile applications increasingly become the primary business interface rather than merely a frontend, organizations will need systematic approaches to resilience validation. The combination of AI-driven test automation with rigorous chaos engineering provides one practical path forward.

\begin{acks}

The authors thank the DragonCrawl team and the uHavoc team for their foundational work and the mobile platform teams for their support in scaling our system.
\end{acks}

\balance
\bibliographystyle{ACM-Reference-Format}
\bibliography{references}

\end{document}